# On Injection in Intrinsic Single-Carrier Devices

Jason A. Röhr[1,2]*

[1]Chemical and Biomolecular Engineering, Tandon School of Engineering, New York University, Brooklyn, New York 11201, USA

[2]Singh Center for Nanotechnology, School of Engineering & Applied Sciences, University of Pennsylvania, Philadelphia, Pennsylvania 19104, USA

*jarohr@seas.upenn.edu

By considering the changes in the interface charge-carrier densities of a single-carrier device as a function of injection-barrier heights and comparing these to the equilibrium, background charge-carrier density of a device with Ohmic contacts, we calculate simple conditions for when these barriers are expected to limit injection and therefore significantly affect space-charge-limited currents in the device. We show that these conditions depend on the device temperature, semiconductor relative permittivity, and effective density of states, but most importantly the thickness of the semiconducting film being probed. This is in accordance with previous observations and similar derived expressions for when defects influence single-carrier devices. The conditions described herein can be used to aid the design of single-carrier devices for space-charge-limited current measurements that are not limited by injection.

## I. INTRODUCTION

Space-charge limited current (SCLC) measurements are commonly used to selectively probe electron or hole charge-transport characteristics of relatively intrinsic semiconductors, via single-carrier devices, and have been used to characterise both organic[1,2], inorganic[3,4], and hybrid semiconductors[5–7]. As SCLC measurements can be highly sensitive to the influence of defects and the choice of contact materials, i.e., whether they are Ohmic or large injection barriers are present, they can in principle be used to not only measure the electron or hole mobilities, $\mu_{n/p}$, but also probe defect characteristics[8] and injection barrier heights, $q\phi$, where $q$ is the elementary charge and $\phi$ is the barrier potential of the non-Ohmic contact[9]. However, to measure these characteristics, sophisticated models, that are more complicated than the most commonly used analytical models, such as the *Mott-Gurney* law[10,11],

$$J = \frac{9}{8}\mu_{n/p}\varepsilon_r\varepsilon_0 \frac{V^2}{L^3} \qquad (1)$$

must typically be employed[8,12]. In **Eq. 1**, $\varepsilon_0$ is the permittivity of free space, $\varepsilon_r$ is the relative permittivity, $V$ is the applied voltage and $L$ is the thickness of the probed semiconductor layer in a sandwich-type single-carrier device. Conversely, single-carrier devices can be designed, and SCLC measurements can be conducted, under conditions where the effects from defects and injection barriers are no longer observable in the current density-voltage (*J-V*) curves, allowing for analysis with simple analytical models and easing the requirements for sophisticated modelling tools if these conditions can be identified and met.

It has been shown that by reducing the thickness of the measured semiconductor that the effect of traps and doping can be neglected due to the increase of the background charge-carrier density that is either compensating for the trapped charge carriers or masking the doping density[13,14]. We have also observed a similar phenomenon pertaining to injection, namely that a large injection barrier will influence thinner devices to a larger degree and that injection barriers are insignificant if the semiconducting layer is made significantly thick[13,15]. This is likely due to $q\phi$ reducing both the background charge-carrier density and limiting injection from the non-Ohmic contact as a voltage is applied across the device. If a condition can be derived for how thick the semiconducting layer must be to render injection barriers insignificant, single-carrier devices can be designed such that barriers can be neglected during the analysis.

By considering the changes in the interface charge-carrier densities as a function of $\phi$ and comparing these to the background charge-carrier density of a device with Ohmic contacts, we here derive simple conditions for when non-Ohmic contacts significantly affect the *J-V* curves obtained from SCLC measurements. We use drift-diffusion simulations to verify the derived conditions and show that they indeed predict when non-Ohmic contacts influence the device. We show that these expressions depend on the device temperature, the semiconductor relative permittivity and effective density of states (DOS), but most importantly the semiconducting film thickness (as this condition scales with $\ln\{L^{-2}\}$). The derived conditions can be used to estimate when injection barriers of finite heights can appear to be perfectly Ohmic and can therefore be used to design intrinsic single-carrier devices that are not affected by injection limitation, improving the ease of analysis.

## II. NUMERICAL DRIFT-DIFFUSION MODEL

We compare the derived analytical expressions with numerical drift-diffusion simulations. This allows us to test their validity over a range of injection barrier heights ($q\phi = 0.0$ eV– $0.6$ eV) and semiconductor thicknesses ($L = 10$ nm–$10$ μm), while ensuring that certain semiconductor characteristics such as traps and doping, can be omitted, and quantities such as the device temperature, $T$, $\varepsilon_r$ and $\mu_{n/p}$ are kept fixed. We have previously used this approach as it allows for an elegant comparison between the derived analytical expressions and drift-diffusion models that has been used to analyse experimental data[6,16–19].

The numerical model used for the drift-diffusion simulations, MacKenzie's *OghmaNano* (previously called *general-purpose photovoltaic device model*)[13,20], solves the drift–diffusion equations for electron and hole current flow, $J_{n/p}$,

$$J_n(x) = qn(x)\mu_n F(x) + qD_n \frac{dn(x)}{dx} \tag{2}$$

$$J_p(x) = qp(x)\mu_p F(x) - qD_p \frac{dp(x)}{dx} \tag{3}$$

along with Poisson's equation describing the electrostatic potential, $\varphi$,

$$\varepsilon_0 \nabla \varepsilon_r \nabla \varphi(x) = -\rho(x) \tag{4}$$

In **Eqs. 2-4**, $n$ and $p$ are the total free electron and hole densities, $D_{n/p}$ are the Einstein–Smoluchowski diffusion coefficients, $F$ is the electric field, $\varphi$ is the electric potential, and $\rho$ is the total charge density. While the drift-diffusion model solves for both electron and hole transport, we are only considering electron-only single-carrier devices for simplicity. Hole current and band-to-band recombination (not included in the equations above) can therefore be neglected during both the drift-diffusion simulations and the analytical derivations due to the very low density of holes in an electron-only device[21]. For this reason, we will only consider the electron mobility, $\mu_n$, during the analytical derivations while setting $\mu_n = \mu_p$ in the numerical model.

The boundary conditions in *OghmaNano* can be set by the interface charge-carrier densities at $x = 0$ (left-hand side) and $x = L$ (right-hand side), namely $n_{\text{int},l}$ and $n_{\text{int},r}$, via the boundary injection-barrier heights, $q\phi_l$ and $q\phi_r$,

$$n_{\text{int},l} = N_C \exp\left(-\frac{q\phi_l}{k_B T}\right) \quad (5)$$

$$n_{\text{int},r} = N_C \exp\left(-\frac{q\phi_r}{k_B T}\right) \quad (6)$$

where $N_C$ is the effective electron DOS. For a symmetric single-carrier device we have that $\phi_l = \phi_r$. Since we will only consider the symmetric case in this study, we will simply refer to the injection barriers collectively as $q\phi$, with their resulting charge-carrier densities written as $n_{\text{int}}$. The single-carrier devices were calculated using device parameters and material constants chosen to represent a generic trap-free semiconductor/insulator as listed in **Table 1**. Just as the case for the mobility, we set $N_C = N_V$ in the numerical model.

**Table 1** – Model device and materials parameters. While the hole mobility, $\mu_p$, and the effective density of valence band states, $N_V$, are not required for the analytical derivations they are explicitly required within the numerical model.

| Device and materials parameters | Parameter symbol | Value and dimensions |
|---|---|---|
| Thickness | $L$ | 10 nm – 10 µm |
| Band gap | $E_g$ | 3 eV |
| Charge-carrier mobility | $\mu_n$ ($\mu_p$) | 1 cm² V⁻¹ s⁻¹ |
| Effective DOS | $N_C$ ($N_V$) | $10^{20}$ cm⁻³ |
| Relative permittivity | $\varepsilon_r$ | 10 |
| Injection barrier heights | $q\phi_l, q\phi_r$ | 0.0 – 0.6 eV |
| Device temperature | $T$ | 300 K |

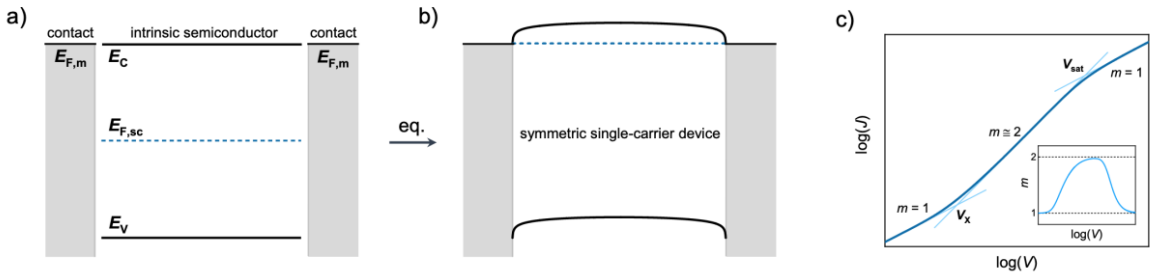

**Figure 1 – a)** Intrinsic semiconductor, prior to thermodynamic equilibrium, with the conduction-band edge aligned with the Fermi level of two metal contacts with equal work functions. **b)** Equilibrium of the metal/semiconductor/metal configuration resulting in a symmetric single-carrier device with two Ohmic injection contacts. The charge-carrier density is now fully dominated by electrons and the current flow across the device will primarily be due to electrons.

**c)** Schematic of an SCLC *J-V* curve, showing the transitions from linear to square-law behaviour back to linear at high voltage. The cross-over voltages between these three regimes ($V_X$ and $V_{sat}$) are depicted. Inset shows the slope of the *J-V* curve on a log-log scale, $m = d\,log(J)/d\,log(V)$, as a function of voltage.

## III. BACKGROUND THEORY

### A. Single-carrier devices

As the name implies, a single-carrier device is a device where only a single charge-carrier type, i.e., either electrons or holes, is responsible for the current under an applied voltage. A symmetric single-carrier device is achieved by matching the work functions, i.e., the contact Fermi levels, $E_{F,m}$, of both contact materials with either the conduction- ($E_C$) or valence-band edge ($E_V$) of the probed semiconductor (**Fig. 1a**), resulting in the device as shown in **Fig. 1b** (after Fermi-level equilibration). Due to the injection of charge-carriers far in excess of the intrinsic charge-carrier density into the semiconductor, the semiconductor mainly contains space charge and the current across the device is therefore space-charge limited[21].

### B. Current density and charge-carrier density

The fact that the current across a single-carrier device is typically space-charge-limited allows for charge-carrier specific charge-transport measurements of the sandwiched semiconductor. In the absence of defects giving rise to traps or doping, energetic disorder and injection barriers, the current density as a function of voltage in a single-carrier device can be described by *de Levie*'s equation at low voltage[22,23], the *Mott-Gurney* law (**Eq. 1**) at intermediate voltage[10,24], and an expression describing saturation current at high voltage, as follows[15,25],

$$J = \begin{cases} 4\pi^2 \dfrac{k_B T}{q} \mu_n \varepsilon_r \varepsilon_0 \dfrac{V}{L^3} & \text{for} \quad 0 < V < V_X \\ \dfrac{9}{8} \mu_n \varepsilon_r \varepsilon_0 \dfrac{V^2}{L^3} & \text{for} \quad V_X < V < V_{sat} \\ q\mu_n N_C \dfrac{V}{L} & \text{for} \quad V_{sat} < V \end{cases} \quad (7)$$

where $k_B T$ is the thermal energy and $V_X$ is the cross-over voltage from the linear regime at low voltage to the Mott-Gurney regime,

$$V_X = \frac{32\pi^2}{9}\frac{k_B T}{q} \tag{8}$$

and $V_{\text{sat}}$ is the cross-over voltage from the Mott-Gurney regime to the linear saturation regime,

$$V_{\text{sat}} = \frac{8}{9}\frac{qL^2 N_C}{\varepsilon_r \varepsilon_0} \tag{9}$$

as shown in **Fig. 1c**. The three models described by **Eq. 7** can, in principle, be used to extract the charge-carrier mobility (here electron mobility) from *J-V* curves obtained from SCLC measurements under the above-mentioned conditions by fitting with the respective models in their appropriate voltage regimes (**Fig. 1c**).

The *de Levie* equation describes the drift current through a single-carrier device at low voltage, where the current flow is only due to the background charge-carrier density that have diffused into the semiconductor during Fermi-level equilibration (hence the inclusion of the *Einstein-Smoluchowski* diffusion coefficient $D = \mu k_B T q^{-1}$), i.e., the charge-carrier density inside the semiconductor before a significant amount of charge carriers are injected due to the applied voltage. Ignoring the intrinsic charge-carrier density, and any effects from defects, i.e., doping and traps, this background charge-carrier density can be expressed as a function of the spatial position inside the semiconductor, $x$, as,

$$n_b(x) = \frac{2\pi^2 \varepsilon_r \varepsilon_0 k_B T}{q^2 L^2}\left[\cos^2\left\{\frac{\pi x}{L} - \frac{\pi}{2}\right\}\right]^{-1}. \tag{10}$$

The above expression is a good description for the charge-carrier density inside the single-carrier device when $V \sim 0$ and no additional injection from the contacts into the semiconductor occurs.

It should here be noted that the expressions for $n_{\text{int}}$ and $n_b$ are not mutually consistent as the value for $n_{\text{int}}$ is always finite for $q\phi \geq 0$ eV; however, $n_b \to \infty$ for $x \to 0$ or $x \to L$. Physically, the charge-carrier density cannot tend to infinity anywhere inside the semiconductor, nor at the interfaces; however, despite $n_b$ not being a good description for the charge-carrier density at the interfaces, it is a useful description in the bulk of the device and will suffice for the present discussion.

When a significant voltage is applied across the device, injection of charge-carriers from the injecting contact, along with the resulting asymmetry in the charge-carrier density can be described by,

$$n_{\text{inj}}(x, V) = \frac{3}{4} \frac{\varepsilon_r \varepsilon_0}{q} \frac{V}{L^{3/2}} x^{-1/2} \qquad (11)$$

giving the total charge-carrier density[13],

$$n_{\text{tot}}(x, V) = n_b(x) + n_{\text{inj}}(x, V). \qquad (12)$$

For $V < V_X$, $n_{\text{inj}}$ can be neglected. In the absence of defects, energetic disorder and injection barriers, the above equation is a useful approximation for the internal charge-carrier density as a function of both $x$ and $V$[13].

Defects in the semiconductor typically lead to either trap sites or doping. On the one hand, traps tend to lower the free charge-carrier density by immobilizing a fraction, where $N_t$ is the density of strap sites. On the other hand, doping tends to increase the free charge-carrier density by introducing free charge carriers from ionized dopants, $n_{\text{dop}}$[13]. In both of these cases, a general condition can be written for how large the trap density or doping density must be before they play a significant role in either lowering or increasing the free charge-carrier density, namely $N_t > \langle n_b \rangle$ and $n_{\text{dop}} > \langle n_b \rangle$[13,14], where $\langle n_b \rangle$ is the harmonic mean of **Eq. 10**[9,13,26],

$$\langle n_b \rangle = \frac{1}{\frac{1}{L} \int_0^L n_b^{-1} \, dx} = \frac{4\pi^2 \varepsilon_r \varepsilon_0 k_B T}{q^2 L^2}. \qquad (13)$$

We now show that similar conditions can be written for the role of the injection barriers and when they significantly affect the current flowing across a single-carrier device.

IV. **RESULTS & DISCUSSION**

A. **Injection-barrier influence on background density**

The difference between the semiconductor conduction-band edge and Fermi level, $\psi(x, V) = E_C(x) - E_{F,\text{sc}}(x, V)$, can generally be described by $-k_B T \ln(n_{\text{tot}} N_C^{-1})$, with $n_{\text{tot}}(x, V)$ given by **Eq. 12**[13]; however, for $V \sim 0$, $n_{\text{inj}}$ can be neglected and $\psi(x)$ can be described by[13],

$$\psi(x) = -k_\mathrm{B}T \ln\left(\frac{2\pi^2 \varepsilon_\mathrm{r}\varepsilon_0 k_\mathrm{B}T}{q^2 L^2 N_\mathrm{C}}\left[\cos^2\left\{\frac{\pi x}{L} - \frac{\pi}{2}\right\}\right]^{-1}\right). \tag{13}$$

**Eq. 13** is a good approximation for $\psi$, as shown by comparing the equation to results from drift-diffusion simulations (dashed lines in **Fig. 2a-d**) regardless of the chosen device thickness in the range investigated here (10 nm to 10 μm), especially in the middle of the single-carrier device ($x = L/2$) for larger thicknesses; however, this is not always the case when injection barriers are present at the interfaces.

From **Fig. 2a-d** it is seen that increasing $q\phi$ has the effect of separating the conduction band edge from the Fermi level across the device (especially at the interfaces). While injection barriers increase the $E_\mathrm{C} - E_\mathrm{F,sc}$ by the same amount regardless of device thickness, a thinner device is influenced to a higher degree due to the very little initial separation between the energy levels. As the charge-carrier density is directly related to $\psi$, a reduction in the background electron density ($n = N_\mathrm{C} \exp[-\psi/k_\mathrm{B}T]$) is generally observed with increased $q\phi$, with thinner devices being affected to a larger degree due to their large densities of background charge-carrier density (**Fig. 2e-h**). In fact, since $n_\mathrm{b} \propto L^{-2}$, it can be observed that as the thickness of the semiconducting layer decreases, so does the influence from the injection barriers. Where even a very small injection barrier of 0.1 eV significantly reduces the electron density in a 10 nm device, the barrier must exceed 0.3 eV in a 10 μm to significantly impact the electron density in the bulk.

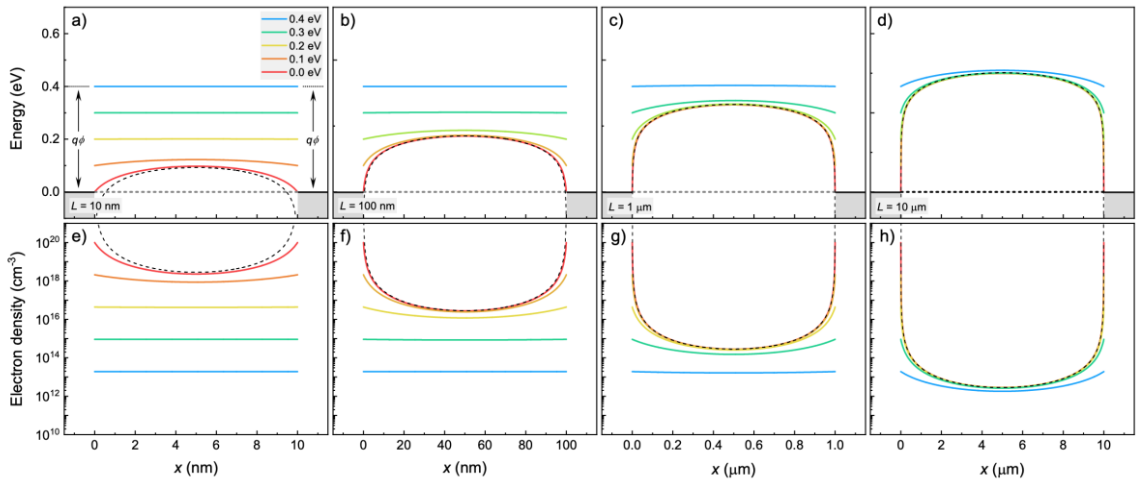

**Figure 2 – a-d)** Calculated $\psi(x)$ curves using drift-diffusion simulations (solid lines) and Eq. 13 (dashed black lines) as a function of semiconductor device thickness (10 nm to 10 μm) and injection barrier heights (0.0 to 0.4 eV). **e-h)** Calculated background charge-carrier density as a function of $x$ using drift-diffusion simulations (solid lines) and Eq. 10 (dashed black lines).

## B. Injection-limited current flow

Even though the trends shown in **Fig. 2** are only for $V = 0$, these translate directly to the cases where current is flowing across the device under a low applied voltage. This is because the current flow in such a low-voltage regime is not primarily due to injected charge-carriers from an applied voltage, but rather due to the background charge-carrier density whose density is significantly affected by the injection barriers.

     **Fig. 3** shows *J-V* curves from simulated, intrinsic single-carrier devices with corresponding changes in device parameters as was used for calculating $\psi$ and the background charge-carrier density in **Fig. 2**. Similar to how the background charge-carrier density was heavily influenced by injection barriers if the device is thin, so is the current density. In **Fig. 3a** it can be seen that while injection barriers indeed limit the current flow at low voltage when significantly large, injection barriers limit current flow at high voltage more readily, with a transition from $J \propto V^2$ to $J \propto V$ and a shift in $V_{\text{sat}}$ to a lower overall value. In fact, if the injection barriers are sufficiently large, $V_{\text{sat}}$ will become smaller than $V_{\text{X}}$ and the current density will be proportional to the applied voltage across the entire voltage range[21]. As the semiconductor thickness is increased, the effect from injection barriers decreases (**Fig. 3b-d**); however, the high-voltage regime is still more readily influenced despite of the device thickness.

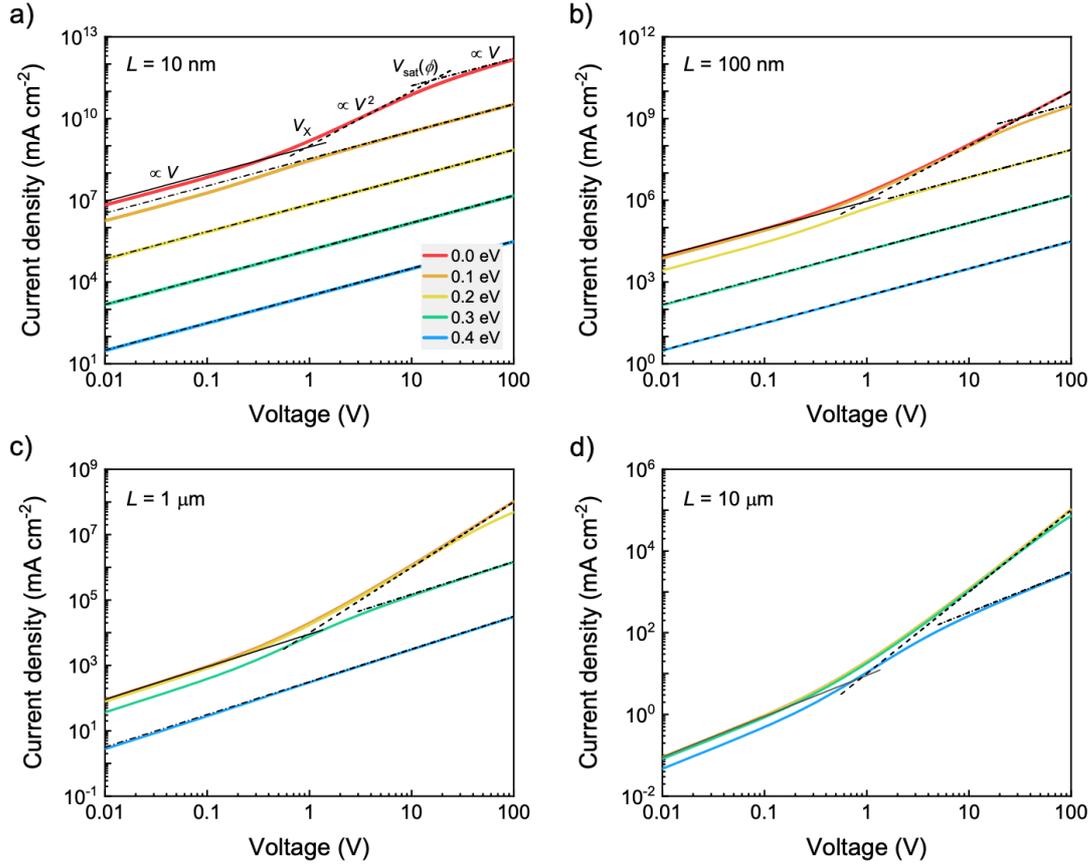

**Figure 3 – a)** *J-V* curves from a simulated 10 nm electron-only device as a function of injection barrier heights, 0.0-0.4 eV. The solid black line corresponds to the *de Levie* equation, the dashed line is the *Mott-Gurney* law, and the dot-dashed line is the equation describing saturation current flow (Eq. 7); the same input parameters were used for the analytical models as were used for the numerical calculations (Table 1). The voltage regimes and cross-over voltages, $V_X$ and $V_{sat}$, are denoted. *J-V* curves from simulated 100 nm-10 μm, all as a function of injection barrier heights, are shown in **b), c)** and **d)**.

The expression for the saturation current in **Eq. 7** can be rewritten to account for injection barriers,[15]

$$J^* = q\mu N_C \frac{V}{L} \exp\left(-\frac{q\phi}{k_B T}\right) \quad (14)$$

which results in a voltage shift in the onset to the saturation current regime given by,

$$V_{sat}^* = \frac{8}{9}\frac{qL^2 N_C}{\varepsilon_r \varepsilon_0} \exp\left(-\frac{q\phi}{k_B T}\right). \quad (15)$$

The onset to saturation will therefore occur at a lower voltage if injection barriers are large. This is especially the case when the semiconducting layer is thin, which is in accordance with the numerical calculations shown in **Fig. 3**. From **Fig. 3** it is also evident that if the influence from injection barriers is sufficiently high, it is no longer possible to analyse the data meaningfully with the analytical models described by **Eq. 7**, in accordance with our previous publication[15].

## C. Derivation of conditions

We here show that conditions for when injection barriers are significant can be derived in two different ways, namely by either observing when the saturation current in the high voltage regime dominates over the square-law behaviour or by considering when the injection barrier heights exceed $\psi$ in the middle of the device at low voltage. While these two conditions will be similar, they signify when injection barriers are significant for the high- and low-voltage regimes, respectfully.

If we define the condition for when the injection barriers are significant in the high-voltage regime as the situation where the current is saturation-dominated before a transition to the *Mott-Gurney* law occurred,

$$V_{\text{sat}}^* < V_X \tag{16}$$

we can then write this simple condition in terms of the injection barrier height as,

$$q\phi > -k_B T \ln\left(\frac{4\pi^2 \varepsilon_r \varepsilon_0 k_B T}{q^2 L^2 N_C}\right). \tag{17}$$

which is equivalent to writing $q\phi > -k_B T \ln(\langle n_b \rangle N_C^{-1})$. The critical injection barrier height, $q\phi_0$, the barrier height that will significantly influence a device of a specific thickness, can be calculated using **Eq. 17** by equating the left- and right-hand side (**Fig. 4a**).

From **Fig. 2** and **3** it can be seen that the current density in the low-voltage regime is not significantly affected until the injection barriers are large enough at the interfaces that they also significantly affect the charge-carrier density in the middle of the device. A second onset expression can therefore be derived by either considering the injection barriers or charge-carrier densities at the contact-semiconductor interfaces (**Eq. 5** and **6**) and comparing these to either $\psi$ in the middle of the device,

$$\psi_0 = \psi(x = L/2) = -k_\mathrm{B}T \ln\left(\frac{2\pi^2 \varepsilon_\mathrm{r}\varepsilon_0 k_\mathrm{B}T}{q^2 L^2 N_\mathrm{C}}\right) \tag{18}$$

or the background charge-carrier density,

$$n_0 = n_\mathrm{b}(x = L/2) = \frac{2\pi^2 \varepsilon_\mathrm{r}\varepsilon_0 k_\mathrm{B}T}{q^2 L^2}. \tag{19}$$

We can now write the condition for when injection barriers are significant as,

$$q\phi > \psi_0 \quad \text{or} \quad n_0 < n_\mathrm{int} \tag{20}$$

Which can be written in terms of the injection barrier height as,

$$q\phi > -k_\mathrm{B}T \ln\left(\frac{2\pi^2 \varepsilon_\mathrm{r}\varepsilon_0 k_\mathrm{B}T}{q^2 L^2 N_\mathrm{C}}\right). \tag{21}$$

**Eq. 21** gives a similar condition to **Eq. 17** for when injection barriers limit current flow inside a single-carrier device at low voltage (**Fig. 4a**); however, **Eq. 21** predicts that the injection barriers have to be slightly higher for a specific thickness than what is predicted from **Eq. 17**, meaning that the current flow at low voltage is influenced less by an injection barrier than a current flow at higher voltage (**Fig. 4a**), as was also shown in **Fig. 3**.

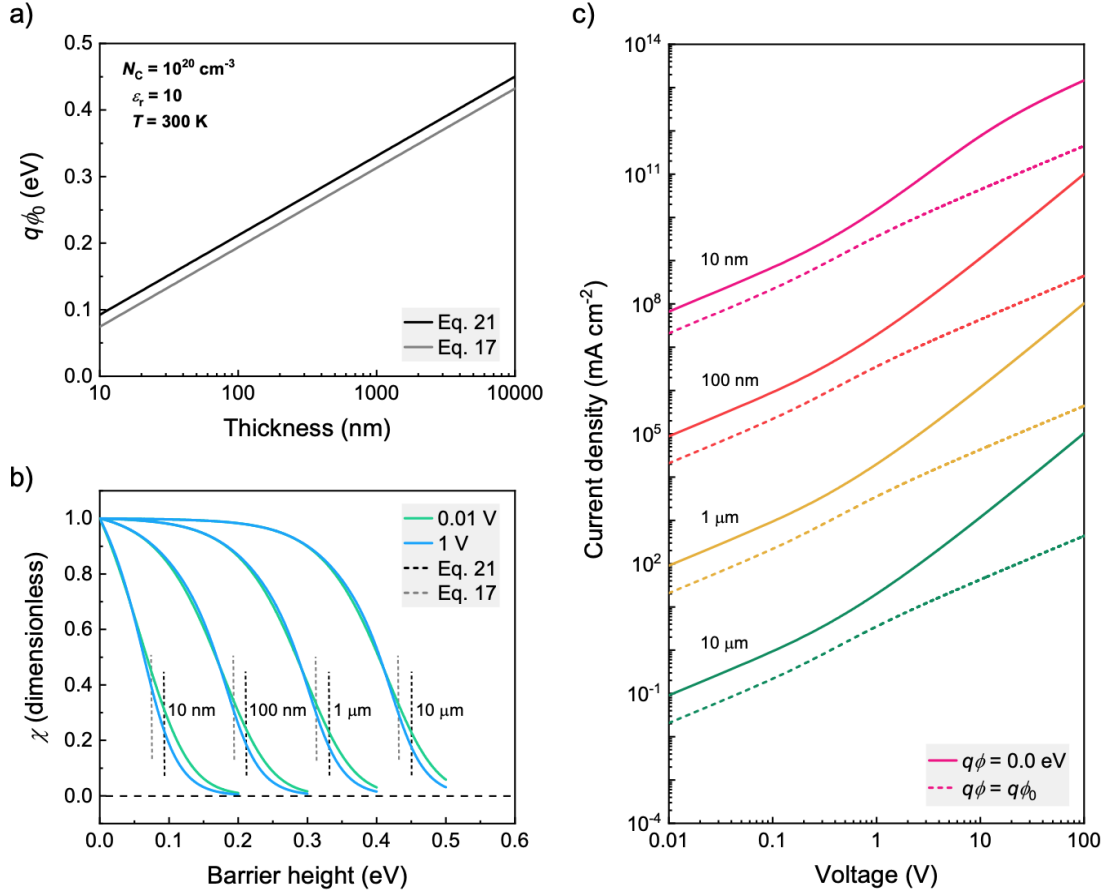

**Figure 4 - a)** Critical injection barrier height as a function of semiconductor thickness calculated using either **Eq. 17** (grey line) or 21 (black line). **b)** $\chi$ as a function of injection barrier height at 0.01 V and 1 V applied voltage for four semiconductor thicknesses ranging from 10 nm to 10 μm. Critical injection barrier heights are shown with dashed lines. **c)** Calculated J-V curves with either Ohmic contacts (solid lines) or with injection barrier heights equal to the critical barrier heights (Eq. 21) at their respective thicknesses (dashed lines).

### D. Verification of conditions

We use the same numerical model used to calculate the energy levels, charge-carrier densities and J-V curves to qualitatively verify **Eq. 17** and **21**. To do so, we define a quantity, $\chi$, given by the ratio of current density for a simulated single-carrier device with injection barriers, $J(\phi > 0)$, to that of a device with Ohmic contacts, i.e., no injection barriers, $J(\phi = 0)$,

$$\chi = \frac{J_{\phi>0}}{J_{\phi=0}}. \tag{22}$$

We evaluate $\chi$ at $V = 0.01$ V and $V = 1$ V, corresponding to situations of low and high applied voltage, as a function of both barrier height at specific device thickness and compare the deviations from unity with **Eq. 17** and **21**.

**Eq. 22** as a function of $q\phi$ for $L$ from 10 nm to 10 μm is shown in **Fig 4b**. Critical barrier heights, as calculated with **Eq. 17** and **21**, are shown as black and grey dashed lines, respectively. It can be seen that $\chi$ approaches 0 at low values for the injection barrier heights (~0.2 eV) for the 10 nm device, regardless of the applied voltage; however, the barrier heights influence the *J-V* curves at 1 V to a higher degree (especially for barrier height values ranging from around 0.06 eV to 0.2 eV) than they do at 0.01 V. As the semiconducting layer increases, so does the transition for when $\chi$ approaches 0, similar to what was observed in **Fig. 3**; this trend follows what is described by **Eq. 17** and **21**, albeit qualitatively. Additionally, *J-V* curves were calculated for devices with thicknesses ranging from 10 nm to 10 μm with either Ohmic contacts or contact barriers equal to $q\phi_0$ as calculated from **Eq. 21** (**Fig. 4c**). A similar trend is observed in all four cases, with the current density approaching a linear *J-V* dependence across all voltages along with a significant decrease in the overall current density, i.e., both a low, intermediate, and high voltage.

### E. Considerations

Whether considering the high- or low-voltage regime, it is clearly observed that measuring single-carrier devices with thicker semiconducting layers can be beneficial as injection barriers will be less significant. The same can be said about materials with high effective DOS, such as organic semiconductors. For example, assuming that an intrinsic, organic semiconductor is being measured, we can then input $\varepsilon_r = 3$, $N \approx 10^{20}$-$10^{21}$ cm$^{-3}$, $T = 300$ K and a semiconductor thickness of 100 nm (as is typical for SCLC measurements on organics), **Eq. 21** predicts that $q\phi$ will have to exceed $\approx$ 0.24-0.30 eV before significantly affecting the current flow. Measuring charge-transport in a 1-10 μm layer of silicon (Si), $\varepsilon_r = 11$, $N \approx 2 \times 10^{19}$ cm$^{-3}$, $q\phi$ would have to exceed 0.29-0.41 eV. Due to the lower effective DOS of Si, even though a much thicker semiconductor layer was measured, the injection barrier tolerance is similar to the organic device. With a proper choice of contact materials, achieving injection barrier heights < 0.3 eV should be possible for many semiconducting materials being probed.

While the above results show that it is indeed beneficial to perform SCLC measurements on intrinsic single-carrier devices with thicker semiconducting layers, as this would allow for analysis with simple analytical models, a dilemma arises if these semiconductors are highly defective or disordered. The thicker the semiconducting layer, the more influenced the SCLC *J-V* curves are from defects and disorder, making it meaningless to analyse the data with the models presented in **Eq. 7**. So, while we can now reasonably identify under what conditions a single factor can be ignored, this interplay between semiconductor thickness, injection barriers and defects highlights the need for models that can account for all these factors explicitly, especially if SCLC is used to probe highly defective or disordered semiconductors. These would have to be either highly sophisticated analytical models yet to be derived or drift-diffusion models such as the one used herein[9,13].

## V. CONCLUSIONS

We derived two conditions for when injection barriers at the interface between the contact and semiconductor in single-carrier devices are expected to limit injection and therefore affect SCLC *J-V* curves at both low and high voltage. We showed that these conditions depend on the device temperature, the relative permittivity and effective DOS of the semiconductor, but most importantly the semiconductor thickness as the conditions are directly related to the equilibrium, background charge-carrier density of a single-carrier device. Our results are in accordance with previous observations showing that thinner devices are more readily influenced by injection barriers, and the derived conditions are similar, albeit showing opposite trends, to previously derived conditions for when traps and doping influence SCLC curves. While our results highlight the intricate interplay between the influence of injection barriers at the interfaces and defects in the bulk of the semiconductor, and hence the need for more sophisticated SCLC models, the conditions derived and discussed herein can be used to design single-carrier devices that are not limited by injection, greatly easing the subsequent analysis of the *J-V* curves by eliminating the need to account for injection barriers.

## ACKNOWLEDGEMENTS

J.A.R. would like to thank Prof. Roderick MacKenzie for his tireless efforts developing the *OghmaNano* software, and Prof. André D. Taylor for his support.